\PassOptionsToPackage{unicode}{hyperref}
\PassOptionsToPackage{hyphens}{url}
\PassOptionsToPackage{dvipsnames,svgnames,x11names}{xcolor}
\documentclass[10pt,a4paper,onecolumn]{article}
\usepackage{marginnote}
\usepackage[rgb,svgnames]{xcolor}
\usepackage{authblk,etoolbox}
\usepackage{titlesec}
\usepackage{calc}
\usepackage{tikz}
\usepackage{hyperref}
\hypersetup{%
    unicode=true,
    pdftitle={FIRM3D: Fast ion reduced models in 3D},
    pdfauthor={Elizabeth Paul, Alexey Knyazev, Michael Czekanski,
Alexandra Lachmann, Abdullah Hyder, Christopher Albert, Matt Landreman},
    colorlinks=true,
    linkcolor=[rgb]{0.0, 0.5, 1.0},
    citecolor=Blue,
    urlcolor=[rgb]{0.0, 0.5, 1.0},
    breaklinks=true
}
\usepackage{caption}
\usepackage{orcidlink}
\usepackage{tcolorbox}
\usepackage{amssymb,amsmath}
\usepackage{ifxetex,ifluatex}
\usepackage{seqsplit}
\usepackage{xstring}

\makeatletter
\@ifundefined{KOMAClassName}{
  \IfFileExists{parskip.sty}{%
    \usepackage{parskip}
  }{
    \setlength{\parindent}{0pt}
    \setlength{\parskip}{6pt plus 2pt minus 1pt}}
}{
  \KOMAoptions{parskip=half}}
\makeatother
\usepackage{graphicx}
\makeatletter
\newsavebox\pandoc@box
\newcommand*\pandocbounded[1]{
  \sbox\pandoc@box{#1}%
  \Gscale@div\@tempa{\textheight}{\dimexpr\ht\pandoc@box+\dp\pandoc@box\relax}%
  \Gscale@div\@tempb{\linewidth}{\wd\pandoc@box}%
  \ifdim\@tempb\p@<\@tempa\p@\let\@tempa\@tempb\fi
  \ifdim\@tempa\p@<\p@\scalebox{\@tempa}{\usebox\pandoc@box}%
  \else\usebox{\pandoc@box}%
  \fi%
}
\def\fps@figure{htbp}
\makeatother
\NewDocumentCommand\citeproctext{}{}
\NewDocumentCommand\citeproc{mm}{%
  \begingroup\def\citeproctext{#2}\cite{#1}\endgroup}
\makeatletter
 \let\@cite@ofmt\@firstofone
 \def\@biblabel#1{}
 \def\@cite#1#2{{#1\if@tempswa , #2\fi}}
\makeatother
\newlength{\cslhangindent}
\newlength{\csllabelwidth}
\newenvironment{CSLReferences}[2] 
 {\begin{list}{}{%
  \setlength{\itemindent}{0pt}
  \setlength{\leftmargin}{0pt}
  \setlength{\parsep}{0pt}
  \ifodd #1
   \setlength{\leftmargin}{\cslhangindent}
   \setlength{\itemindent}{-1\cslhangindent}
  \fi
  \setlength{\itemsep}{#2\baselineskip}}}
 {\end{list}}
\usepackage{calc}

\ifLuaTeX
\usepackage[bidi=basic,shorthands=off,]{babel}
\else
\usepackage[bidi=default,shorthands=off,]{babel}
\fi
\ifLuaTeX
  \usepackage{selnolig} 
\fi
\providecommand{\tightlist}{%
  \setlength{\itemsep}{0pt}\setlength{\parskip}{0pt}}

\usepackage{float}
\let\origfigure\figure
\let\endorigfigure\endfigure
\renewenvironment{figure}[1][2] {
    \expandafter\origfigure\expandafter[H]
} {
    \endorigfigure
}

\usepackage[top=3.5cm, bottom=3cm, right=1.5cm, left=1.0cm,
            headheight=2.2cm, reversemp, includemp, marginparwidth=4.5cm]{geometry}

\titleformat{\section}
  {\normalfont\sffamily\Large\bfseries}
  {}{0pt}{}
\titleformat{\subsection}
  {\normalfont\sffamily\large\bfseries}
  {}{0pt}{}
\titleformat{\subsubsection}
  {\normalfont\sffamily\bfseries}
  {}{0pt}{}
\titleformat*{\paragraph}
  {\sffamily\normalsize}

\usepackage{fancyhdr}
\makeatletter
\let\ps@plain\ps@fancy
\fancyheadoffset[L]{4.5cm}
\fancyfootoffset[L]{4.5cm}

\definecolor{linky}{rgb}{0.0, 0.5, 1.0}

\newtcolorbox{repobox}
   {colback=red, colframe=red!75!black,
     boxrule=0.5pt, arc=2pt, left=6pt, right=6pt, top=3pt, bottom=3pt}

\newcommand{\ExternalLink}{%
   \tikz[x=1.2ex, y=1.2ex, baseline=-0.05ex]{%
       \begin{scope}[x=1ex, y=1ex]
           \clip (-0.1,-0.1)
               --++ (-0, 1.2)
               --++ (0.6, 0)
               --++ (0, -0.6)
               --++ (0.6, 0)
               --++ (0, -1);
           \path[draw,
               line width = 0.5,
               rounded corners=0.5]
               (0,0) rectangle (1,1);
       \end{scope}
       \path[draw, line width = 0.5] (0.5, 0.5)
           -- (1, 1);
       \path[draw, line width = 0.5] (0.6, 1)
           -- (1, 1) -- (1, 0.6);
       }
   }

\definecolor{c53baa1}{RGB}{83,186,161}
\definecolor{c202826}{RGB}{32,40,38}

\patchcmd{\@maketitle}{center}{flushleft}{}{}
\patchcmd{\@maketitle}{center}{flushleft}{}{}
\patchcmd{\@maketitle}{\LARGE}{\LARGE\sffamily}{}{}
\def\maketitle{{%
  
  \AB@maketitle}}
\renewcommand\AB@affilsepx{ \protect\Affilfont}
\renewcommand\AB@affilnote[1]{{\bfseries #1}\hspace{3pt}}
\renewcommand{\affil}[2][]%
   {\newaffiltrue\let\AB@blk@and\AB@pand
      \if\relax#1\relax\def\AB@note{\AB@thenote}\else\def\AB@note{#1}%
        \setcounter{Maxaffil}{0}\fi
        \begingroup
        \let\href=\href@Orig
        \let\protect\@unexpandable@protect
        \def\thanks{\protect\thanks}\def\footnote{\protect\footnote}%
        \@temptokena=\expandafter{\AB@authors}%
        {\def\\{\protect\\\protect\Affilfont}\xdef\AB@temp{#2}}%
         \xdef\AB@authors{\the\@temptokena\AB@las\AB@au@str
         \protect\\[\affilsep]\protect\Affilfont\AB@temp}%
         \gdef\AB@las{}\gdef\AB@au@str{}%
        {\def\\{, \ignorespaces}\xdef\AB@temp{#2}}%
        \@temptokena=\expandafter{\AB@affillist}%
        \xdef\AB@affillist{\the\@temptokena \AB@affilsep
          \AB@affilnote{\AB@note}\protect\Affilfont\AB@temp}%
      \endgroup
       \let\AB@affilsep\AB@affilsepx
}
\makeatother

\renewcommand\Affilfont{\sffamily\small\mdseries}
\usepackage{fontspec}
\defaultfontfeatures{Scale=MatchLowercase}
\defaultfontfeatures[\sffamily]{Ligatures=TeX}
\defaultfontfeatures[\rmfamily]{Ligatures=TeX,Scale=1}

\ifPDFTeX\else
\fi
\IfFileExists{upquote.sty}{\usepackage{upquote}}{}
\IfFileExists{microtype.sty}{
  \usepackage[]{microtype}
  \UseMicrotypeSet[protrusion]{basicmath} 
}{}

\usepackage{fontsetup} 

\PassOptionsToPackage{usenames,dvipsnames}{color} 
\ifLuaTeX
  \usepackage{selnolig}  
\fi

\title{FIRM3D: Fast ion reduced models in 3D}

\author[1%
\ensuremath\mathparagraph]{Elizabeth Paul%
  \,\orcidlink{0000-0002-9355-5595}\,%
}
\author[1%
]{Alexey Knyazev%
  \,\orcidlink{0000-0001-8333-859X}\,%
}
\author[2%
]{Michael Czekanski%
  \,\orcidlink{0009-0005-2520-3415}\,%
}
\author[1%
]{Alexandra Lachmann%
  \,\orcidlink{0000-0002-8341-107X}\,%
}
\author[1%
]{Abdullah Hyder%
  \,\orcidlink{0000-0003-4410-3661}\,%
}
\author[3%
]{Christopher Albert%
  \,\orcidlink{0000-0003-4773-416X}\,%
}
\author[4%
]{Matt Landreman%
  \,\orcidlink{0000-0002-7233-577X}\,%
}

\affil[1]{Department of Applied Physics and Applied Mathematics,
Columbia University, USA%
}
\affil[2]{Department of Statistics and Data Science, Cornell University,
USA%
}
\affil[3]{Graz University of Technology, Austria%
}
\affil[4]{University of Maryland, College Park, USA%
}
\affil[$\mathparagraph$]{Corresponding author}
\date{\vspace{-2.5ex}}

\begin{document}
\maketitle

\marginpar{

  \begin{flushleft}
  \sffamily\small

  {\bfseries DOI:} \href{https://doi.org/N/A}{\color{linky}{N/A}}

  \vspace{2mm}
    {\bfseries Software}
  \begin{itemize}
    \setlength\itemsep{0em}
    \item \href{https://github.com/openjournals}{\color{linky}{Review}} \ExternalLink
    \item \href{https://github.com/openjournals}{\color{linky}{Repository}} \ExternalLink
    \item \href{https://doi.org/10.5281}{\color{linky}{Archive}} \ExternalLink
  \end{itemize}

  \vspace{2mm}
  
    \par\noindent\hrulefill\par

  \vspace{2mm}

  {\bfseries Editor:} \href{https://joss.theoj.org}{Open
Journals} \ExternalLink
   \\
  \vspace{1mm}
    {\bfseries Reviewers:}
  \begin{itemize}
  \setlength\itemsep{0em}
    \item \href{https://github.com/openjournals}{@openjournals}
    \end{itemize}
    \vspace{2mm}
  
    {\bfseries Submitted:} 01 January 1970\\
    {\bfseries Published:} 01 January 1970

  \vspace{2mm}
  {\bfseries License}\\
  Authors of papers retain copyright and release the work under a Creative Commons Attribution 4.0 International License (\href{https://creativecommons.org/licenses/by/4.0/}{\color{linky}{CC BY 4.0}}).

  \end{flushleft}
}

\section{Summary}\label{summary}

The dynamics of energetic particle (EP) species, born from fusion
reactions or plasma heating schemes, are critical for predicting the
behavior of magnetic confinement fusion experiments and future fusion
reactors. Because energetic particles are largely collisionless, the
orbits of Monte Carlo samples drawn from a given distribution function
can be efficiently integrated in prescribed electromagnetic fields. In
addition to the static magneto-hydrodynamic (MHD) equilibrium fields
produced by the electromagnetic coils of a fusion device, MHD waves can
be excited by---and resonantly transport---energetic particle
populations.

FIRM3D is an open-source Python/C++/CUDA software suite for modeling
energetic particle dynamics in 3D magnetic fields, available at
\url{https://github.com/ColumbiaStellaratorTheory/firm3d}. The core guiding-center
integration routines grew out of SIMSOPT
(\citeproc{ref-2021Landreman}{Landreman et al., 2021}), but have been
extended to include additional physics and diagnostics not typically
required in the stellarator optimization context. This standalone
framework enables focused development of energetic particle physics
capabilities with minimal dependencies, making it accessible to the
broader stellarator and plasma physics community.

Components of FIRM3D include:

\begin{itemize}
\tightlist
\item
  Interfaces with MHD equilibrium and wave stability software
  (BOOZ\_XFORM, AE3D, FAR3D).
\item
  CPU and GPU parallelized integration of the guiding center orbit
  equation, with symplectic and Runge-Kutta integrator options.
\item
  Orbit visualization and transport diagnostics, including Poincaré
  maps, orbit classification, and weighted Birkhoff averaging.
\end{itemize}

\section{Statement of need}\label{statement-of-need}

Recent advances in stellarator optimization
(\citeproc{ref-2022LandremanOpt}{Landreman et al., 2022}) have produced
equilibria that satisfy many physics and engineering constraints for a
fusion reactor. One critical feature is the ability to confine guiding
center trajectories of energetic particle species. Through
quasisymmetry---a hidden symmetry of the field strength that provides
integrability of guiding center motion---stellarator magnetic fields can
now be designed with excellent energetic particle confinement.

However, perturbing electromagnetic fields can still transport energetic
particles. The class of MHD waves of primary concern are Alfvén
eigenmodes (AEs), which are driven unstable by free energy in the EP
distribution function and can resonantly transport EPs. Alfvénic
activity is considered the major limitation to alpha confinement in
burning tokamak plasmas (\citeproc{ref-2014Gorelenkov}{Gorelenkov et
al., 2014}), and has been observed in numerous stellarator experiments
(\citeproc{ref-2020Rahbarnia}{Rahbarnia et al., 2020};
\citeproc{ref-2011Toi}{Toi et al., 2011}). Given the recent growth of
the private fusion industry, companies pursuing the stellarator path to
fusion require tools for assessing EP-driven wave stability and the
resulting EP transport.

FIRM3D addresses this need by providing a modular, GPU-accelerated
energetic particle simulation framework that bridges MHD equilibrium
solvers and wave stability codes. It is intended for plasma physicists
and fusion engineers who require EP simulation capabilities beyond those
available in optimization-focused codes. The FIRM3D routines and their
SIMSOPT precursors have already been used in published research: a
survey of EP loss mechanisms (\citeproc{ref-2022Paul}{Paul et al.,
2022}), AE-induced transport analysis
(\citeproc{ref-knyazev2026shear}{Knyazev et al., 2026};
\citeproc{ref-2023Paul}{Paul et al., 2023}), trapped EP resonances
theory (\citeproc{ref-2025Chambliss}{Chambliss et al., 2025}), and
analysis of the Helios alpha confinement
(\citeproc{ref-von2026alpha}{Linden et al., 2026};
\citeproc{ref-swanson2025overview}{Swanson et al., 2025}).

Existing tools such as ASCOT5 (\citeproc{ref-varje2019ascot5}{Varje et
al., 2019}) provide high-fidelity Monte Carlo EP tracking, primarily for
tokamak geometry. FIRM3D complements these by offering tight integration
with the stellarator optimization ecosystem (SIMSOPT, BOOZ\_XFORM),
symplectic and Runge-Kutta integrator options, and native GPU
acceleration via CUDA, in a lightweight open-source Python package.

\section{Structure and capabilities}\label{structure-and-capabilities}

Integration of guiding center trajectories is performed given the
magnetic field, particle initial conditions, and integrator
specification. The equilibrium magnetic field is typically provided
through an interface with BOOZ\_XFORM
(\citeproc{ref-booz_xform}{Landreman, 2021}) via the
\texttt{BoozerMagneticField} class. Since BOOZ\_XFORM computes the
Fourier harmonics of the magnetic field on a uniform radial grid,
Lagrange interpolation is used to evaluate the field throughout the
volume. The magnetic field perturbation corresponding to an MHD mode
from AE3D (\citeproc{ref-2010Spong}{Spong et al., 2010}) or FAR3D
(\citeproc{ref-2024Varela}{Varela et al., 2024}) can then be
superimposed on the interpolated equilibrium field; such modes are
stored on a radial grid of Fourier harmonics in Boozer coordinates.
Helper functions are provided to generate particle initial conditions
from a known distribution function or by preserving a conserved quantity
such as energy or canonical momentum.

Three integrators are available. An interface to the Boost Runge-Kutta
Dormand-Prince 5 method
(\citeproc{ref-BoostOdeint}{Boost.Numeric.Odeint, 2025}) provides
adaptive step size control and dense output. A custom Dormand-Prince 5
implementation adds minimum step size control based on
(\citeproc{ref-2007Press}{Press, 2007}) to prevent excessively small
steps. A symplectic integrator for non-canonical guiding-center orbits
uses the explicit-implicit Euler scheme of
(\citeproc{ref-2020Albert}{Albert et al., 2020}).

Since the performance bottlenecks are field interpolation and trajectory
integration, the Lagrange interpolating polynomials and integrators are
implemented in C++, with Python interfaces via pybind11
(\citeproc{ref-pybind11}{Jakob et al., 2017}). MPI parallelization over
Fourier harmonics and OpenMP parallelization over interpolant nodes
accelerate field setup. Because guiding center trajectories are
independent, Monte Carlo samples are trivially parallelized over CPUs or
GPUs; CUDA kernels implement field interpolation and trajectory
integration on GPU.

Given trajectory data, transport diagnostics include Poincaré plots,
characteristic orbit frequencies, weighted Birkhoff averaging
(\citeproc{ref-duignan2023distinguishing}{Duignan \& Meiss, 2023}), and
orbit classification (\citeproc{ref-2023Albert}{Albert et al., 2023};
\citeproc{ref-2022Paul}{Paul et al., 2022}). Examples of these
capabilities are highlighted below.

Installation instructions and API documentation are available at
\url{https://firm3d.readthedocs.io/}. Examples are available in the
repository at \url{https://github.com/ColumbiaStellaratorTheory/firm3d}.
FIRM3D is released under the MIT License. A suite of unit and regression
tests is run automatically on CPUs and GPUs via continuous integration
on GitHub Actions.

\section{Conservation properties}\label{conservation-properties}

\begin{figure}
\centering
\pandocbounded{\includegraphics[keepaspectratio]{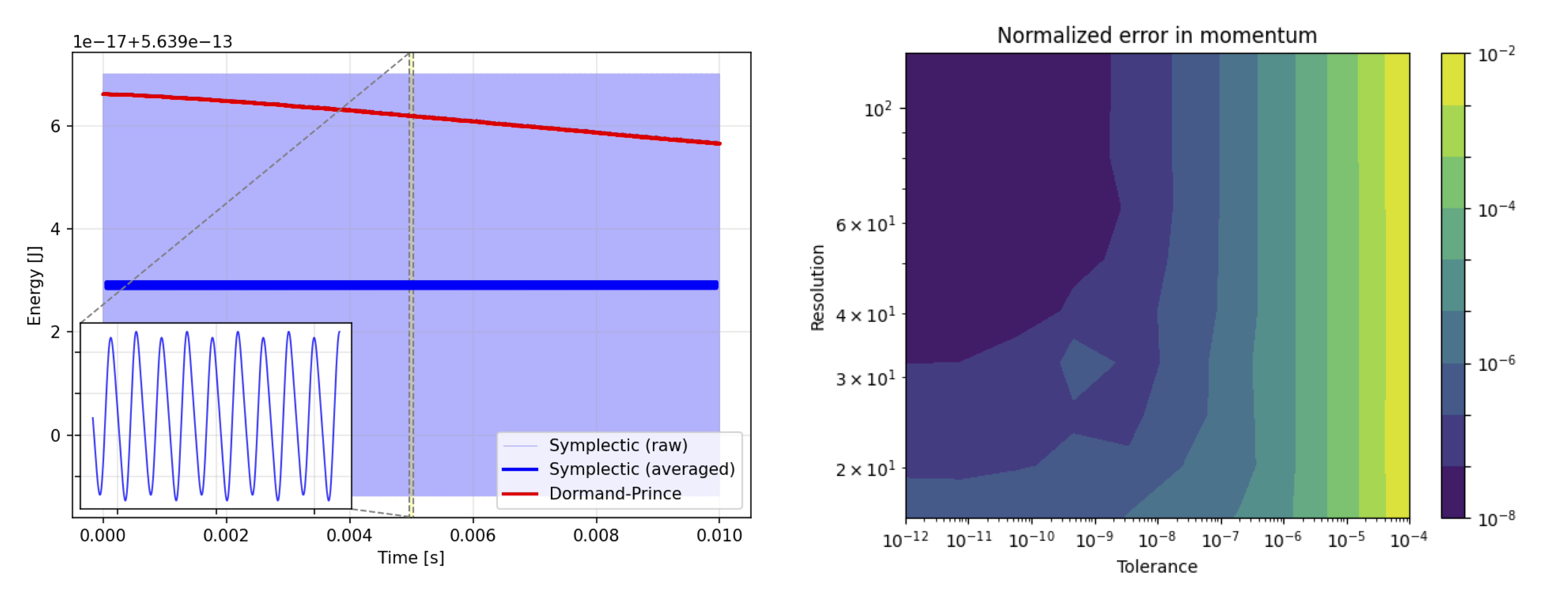}}
\caption{Left: Energy as a function of time for an alpha particle in the
\(\beta = 2.5\%\) Landreman QH configuration. The Dormand-Prince
algorithm exhibits a net energy drift over time, while the symplectic
algorithm exhibits a stable moving time average of the energy (over
\(10^{-4}\) seconds). Right: Relative error in canonical momentum
\(P_{\eta}\) conservation for a perfectly quasisymmetric field. 10
particles are traced in the same configuration for \(10^{-4}\) seconds,
and the maximum error over the trajectory for each particle is computed.
The maximum error over the 10 particles is reported. The
non-quasisymmetric field-strength harmonics are artificially removed so
that momentum conservation is expected. \label{fig:momentum_error}}
\end{figure}

FIRM3D is verified against known conservation laws. For time-independent
fields, the guiding-center Lagrangian conserves total energy \(E\).
Runge-Kutta methods suffer from net energy drift over time, while the
symplectic integrator exhibits long-time stability with a conserved
time-averaged energy, as shown in \autoref{fig:momentum_error}. For a
perfectly quasisymmetric field, the toroidal canonical momentum
\(P_{\eta}\) is also conserved (\citeproc{ref-1995Boozer}{Boozer,
1995}). \autoref{fig:momentum_error} shows the relative error in
\(P_{\eta}\) as a function of the number of Lagrange interpolation nodes
and the Dormand-Prince tolerance. At a tolerance of \(10^{-9}\) and 64
interpolation nodes, the relative error converges to approximately
\(10^{-8}\).

\section{Cross-code comparison}\label{cross-code-comparison}

\begin{figure}
\centering
\pandocbounded{\includegraphics[keepaspectratio]{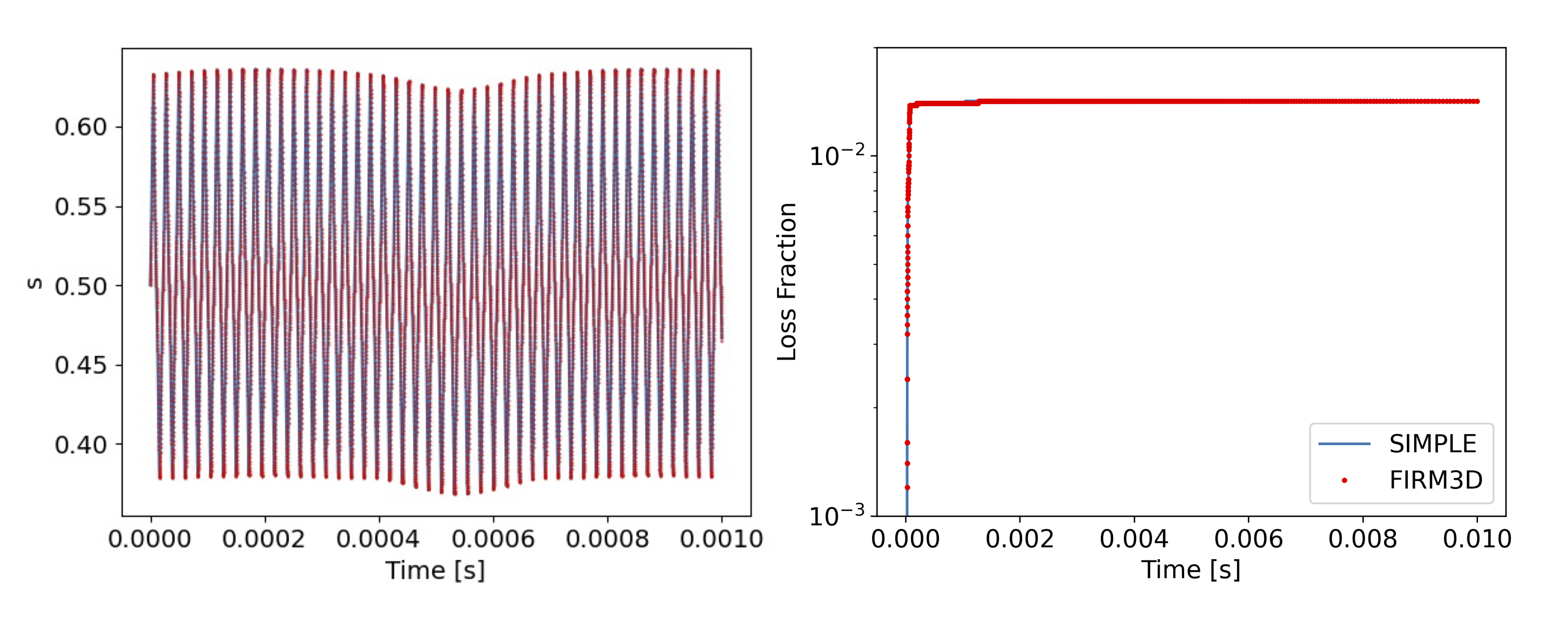}}
\caption{Left: Comparison of a trapped 3.5 MeV alpha particle orbit in
the precise QH equilibrium. The difference in the \(s\) coordinate
between the trajectories at \(10^{-3}\) seconds is
\(7.8\times 10^{-3}\). Right: Comparison of loss fraction for the
precise QA equilibrium. 5000 3.5 MeV alpha particles are sampled from a
fusion birth distribution function and traced for \(10^{-2}\) seconds.
The two codes report identical loss fractions at the end of the
simulation. \label{fig:simple_orbit}}
\end{figure}

\autoref{fig:simple_orbit} shows a benchmark against SIMPLE
(\citeproc{ref-2020Albert}{Albert et al., 2020}), which integrates the
guiding center equations using a symplectic method. We first compare a
trapped 3.5 MeV alpha particle trajectory in the precise QH equilibrium
(\citeproc{ref-2022LandremanPrecise}{Landreman \& Paul, 2022}); the
comparison uses a relatively integrable trajectory since phase-space
chaos generally precludes point-wise agreement between integrators.
FIRM3D used the Dormand-Prince algorithm with relative tolerance
\(10^{-10}\) and 96 Lagrange nodes; SIMPLE used the symplectic Euler
method with 4096 timesteps per toroidal transit. The relative error in
the \(s\) coordinate at \(10^{-3}\) seconds is \(7.8\times 10^{-3}\). We
next compare loss fractions for 5000 3.5 MeV alpha particles sampled
from a fusion birth distribution in the precise QA equilibrium
(\citeproc{ref-2022LandremanPrecise}{Landreman \& Paul, 2022}) traced
for \(10^{-2}\) seconds; the two codes report identical loss fractions.

\section{Scaling on GPUs and CPUs}\label{scaling-on-gpus-and-cpus}

\begin{figure}
\centering
\includegraphics[width=\linewidth,height=5cm,keepaspectratio]{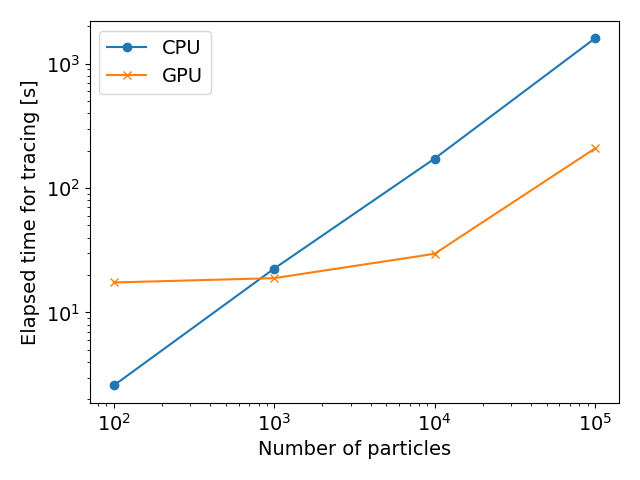}
\caption{Scaling of tracing a fusion birth distribution in the Wistell-A
equilibrium on 1 Perlmutter CPU node (128 CPU threads) and 1 NVIDIA
A100, as a function of number of Monte Carlo samples.
\label{fig:cpu_scaling}}
\end{figure}

\autoref{fig:cpu_scaling} shows wall-clock scaling on the NERSC
Perlmutter cluster. Particles are sampled from a fusion birth
distribution in the Wistell-A equilibrium and integrated for \(10^{-2}\)
seconds. For fewer than \(10^3\) samples the CPU calculation is more
efficient due to GPU launch latency; above \(10^3\) samples the GPU
calculation is approximately an order of magnitude faster. Details of
the GPU implementation are described in
(\citeproc{ref-2026Czekanski}{Czekanski et al., 2026}).

\section{Example applications}\label{example-applications}

\begin{figure}
\centering
\pandocbounded{\includegraphics[keepaspectratio]{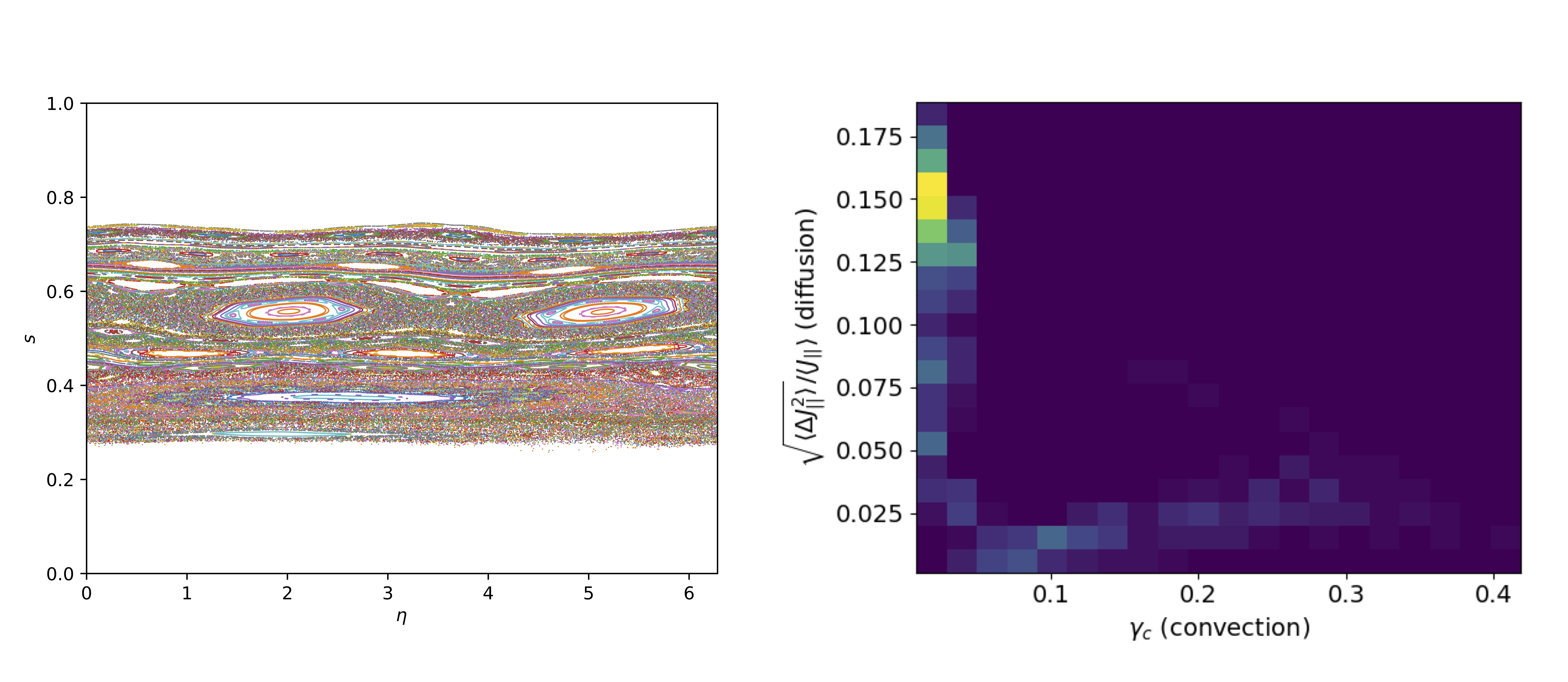}}
\caption{Left: Trapped particle Poincaré map showing chaotic layers
responsible for banana-drift diffusion. Right: Measures of convective
and diffusive transport indicate banana-trapped orbits undergo banana
diffusion. \label{fig:classification}}
\end{figure}

\begin{figure}
\centering
\includegraphics[width=0.6\linewidth,height=\textheight,keepaspectratio]{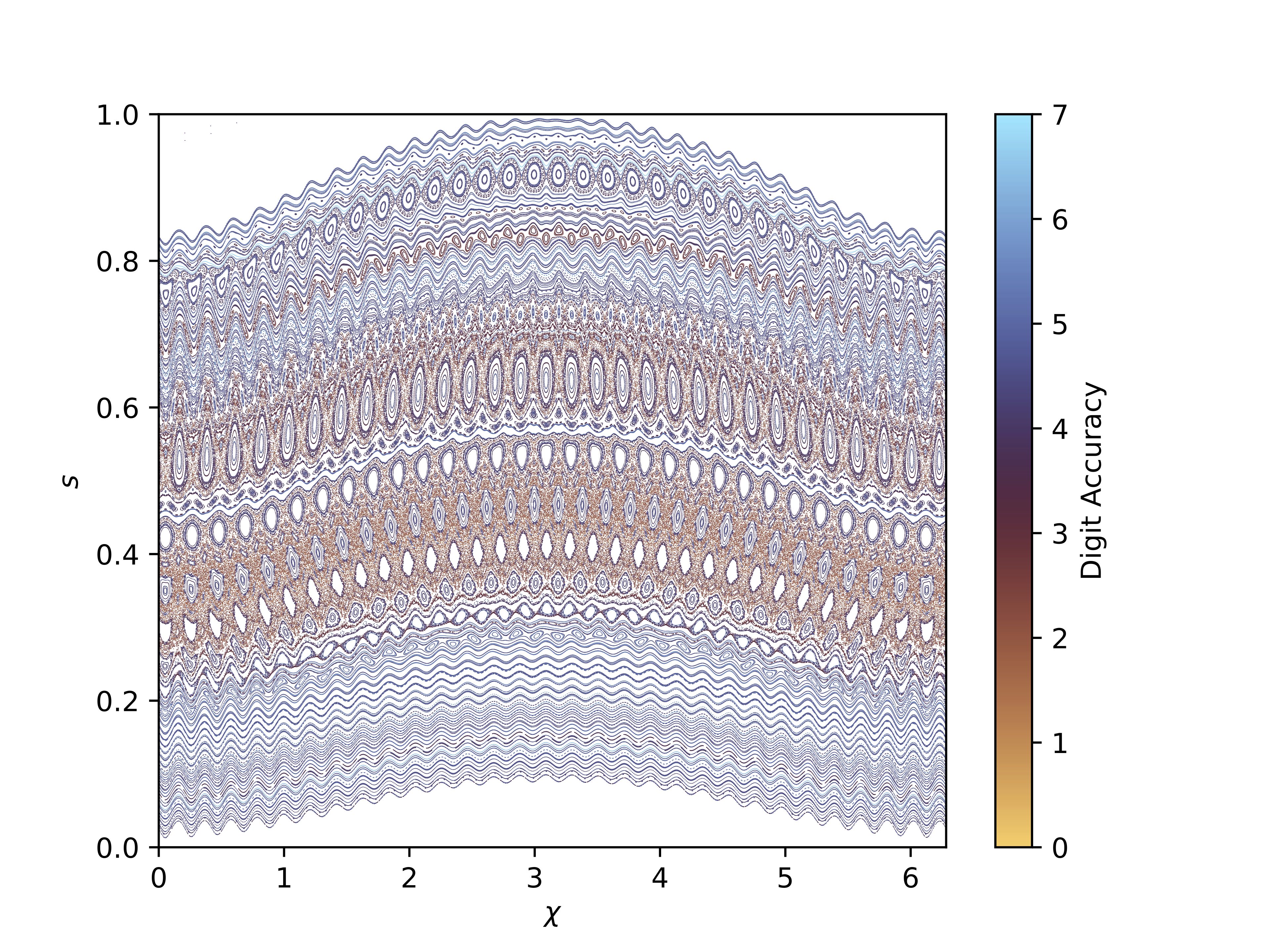}
\caption{A kinetic Poincaré plot colored by the digit accuracy of the
weighted Birkhoff average, an indicator of chaos. \label{fig:wba}}
\end{figure}

FIRM3D's orbit classification, transport diagnostics, and AE-induced
transport capabilities are illustrated in \autoref{fig:classification}
and \autoref{fig:wba}. For the former, \(5\times 10^{5}\) alpha
particles are sampled from a fusion birth distribution in the
\(\beta = 2.5\%\) QA configuration
(\citeproc{ref-2022LandremanPrecise}{Landreman \& Paul, 2022}) and
traced for \(10^{-2}\) seconds (0.62\% lost). Trapping class (banana,
ripple, barely trapped) is identified along each trajectory; of the lost
particles, 29\% are banana-class, 0.12\% ripple-trapped, 0.03\% barely
trapped, and 4.8\% transition between classes, with 66\% exiting
promptly before classification is possible. The normalized variation
\(\sqrt{\langle \Delta J_{\|}^2 \rangle}/\langle J_{\|} \rangle\) of the
parallel adiabatic invariant \(J_{\|} = \oint dl\, v_{\|}\) measures
integrability (\citeproc{ref-2023Albert}{Albert et al., 2023}), and
\(\gamma_c\) measures convective transport (\citeproc{ref-2022Paul}{Paul
et al., 2022}); most lost banana particles exceed the 1\%
non-integrability threshold while \(\gamma_c < 0.2\), indicating orbital
chaos drives the losses. For \autoref{fig:wba}, passing particles with
\(\mu/E = 0.1\) are traced in the \(\beta = 2.5\%\) QA equilibrium
(\citeproc{ref-2022LandremanOpt}{Landreman et al., 2022}) with a
single-harmonic AE (\(m = 30\), \(n = 14\), \(\omega = 136\) kHz; Table
1 of (\citeproc{ref-2023Paul}{Paul et al., 2023})). The weighted
Birkhoff average of \(P_{\zeta}\)
(\citeproc{ref-duignan2023distinguishing}{Duignan \& Meiss, 2023})
serves as an integrability diagnostic, with digit accuracy below 3
indicating chaotic motion (\citeproc{ref-knyazev2026shear}{Knyazev et
al., 2026}).

\section{Acknowledgements}\label{acknowledgements}

We acknowledge the SIMSOPT development team for providing the
foundational guiding center integration routines. We acknowledge funding
through the U.S. Department of Energy under contracts DE-SC0024630,
DE-SC0024548, and DE-AC02-09CH11466, and through the Simons Foundation
collaboration `Hidden Symmetries and Fusion Energy,' Grant No.~601958.
This research used resources of the National Energy Research Scientific
Computing Center (NERSC), a DOE Office of Science User Facility, under
NERSC award ERCAP0031926.

\section*{References}\label{references}
\addcontentsline{toc}{section}{References}

\protect\phantomsection\label{refs}

\end{document}